\documentclass[aps,prx,twocolumn, superscriptaddress,amsmath,amssymb,longbibliography]{revtex4-2}
\usepackage{xcolor}
\usepackage{amsmath}
\usepackage{graphicx}% Include figure files
\usepackage{dcolumn}% Align table columns on decimal point
\definecolor{webblue}{HTML}{1111aa}
\usepackage[
colorlinks=true,
linkcolor=webblue,
citecolor=webblue,
urlcolor=webblue,
]{hyperref}
\usepackage{bm}% bold math
\usepackage[T1]{fontenc}
\usepackage{hyphenat}

\begin{document}

\preprint{APS/123-QED}

\title{\textit{p}-wave orbital angular momentum texture in a chiral crystal}

\author{Dongjin Oh}
\email[Corresponding Author:$~$]{djeeoh@gmail.com}
\affiliation{Department of Physics, Massachusetts Institute of Technology, Cambridge, MA, USA}

\author{Chiara Pacella}
\email[Corresponding Author:$~$]{chiara.pacella2@unibo.it}
\affiliation{Department of Physics and Astronomy, University of Bologna, Bologna, Italy}
\affiliation{Max Planck Institute for the Structure and Dynamics of Matter, Hamburg, Germany}

\author{Xiangyu Luo}
\affiliation{Department of Physics, Massachusetts Institute of Technology, Cambridge, MA, USA}

\author{Chris Jozwiak}
\affiliation{Advanced Light Source, Lawrence Berkeley National Laboratory, Berkeley, CA, USA}

\author{Aaron Bostwick}
\affiliation{Advanced Light Source, Lawrence Berkeley National Laboratory, Berkeley, CA, USA}

\author{Eli Rotenberg}
\affiliation{Advanced Light Source, Lawrence Berkeley National Laboratory, Berkeley, CA, USA}

\author{Mats Leandersson}
\affiliation{Max IV Laboratory, Lund University, Lund, Sweden}

\author{Craig Polley}
\affiliation{Max IV Laboratory, Lund University, Lund, Sweden}

\author{Angel Rubio}
\affiliation{Max Planck Institute for the Structure and Dynamics of Matter, Hamburg, Germany}
\affiliation{Nano-Bio Spectroscopy Group, Departmento de Física de Materiales, Universidad del País Vasco, San Sebastián, Spain}
\affiliation{Center for Computational Quantum Physics, The Flatiron Institute, New York, NY, USA}

\author{Domenico Di Sante}
\email[Corresponding Author:$~$]{domenico.disante@unibo.it}
\affiliation{Department of Physics and Astronomy, University of Bologna, Bologna, Italy}

\author{Riccardo Comin}
\email[Corresponding Author:$~$]{rcomin@mit.edu}
\affiliation{Department of Physics, Massachusetts Institute of Technology, Cambridge, MA, USA}

\begin{abstract}
The spin and orbital angular momentum (SAM and OAM) are conceptually analogous, yet their roles in condensed matter systems have not been often treated on equal footing. While SAM has been extensively explored, OAM has long been regarded as quenched in crystalline environments and thus largely overlooked. Recent experimental and theoretical advances, however, have demonstrated that OAM can drive a variety of novel electronic phenomena, highlighting the importance of probing OAM textures in the electronic band structure. Here, we investigate the momentum-space OAM texture of (TaSe$_{4}$)$_{2}$I, a one-dimensional chiral crystal. Using circular-dichroism angle-resolved photoemission spectroscopy (CD-ARPES), we uncover a $p$-wave OAM texture accompanied by OAM dipole structures. This orbital $p$-wave texture is intimately connected to, and thus controllable by, the chirality of the host lattice. Complementary spin-resolved ARPES measurements and first-principles calculations reveal that the OAM polarization overwhelmingly dominates the low-energy electronic properties of (TaSe$_{4}$)$_{2}$I, far exceeding the SAM polarization. These observations represent the experimental verification of a new type of OAM texture in crystalline materials. Most importantly, these findings underscore a promising material platform for spinless orbitronics applications and lay the foundation for realizing multipolar OAM textures$-$orbital counterparts of the spin textures in unconventional magnets.

\end{abstract}

\maketitle

\section{Introduction}
\begin{figure*}[htbp!]
	\includegraphics[width=16.5cm]{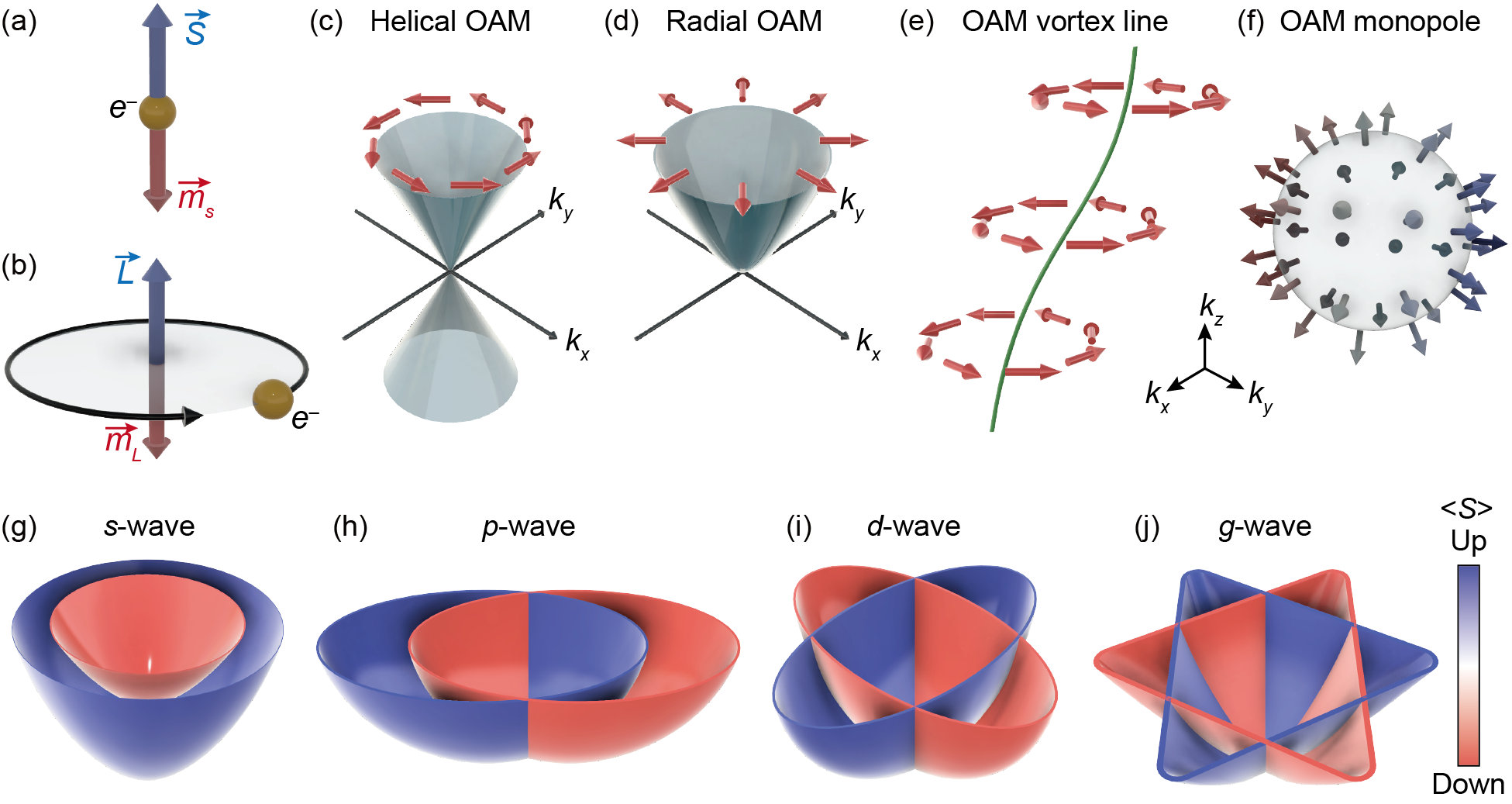}
	\caption{
        $\vec{k}$-space SAM and OAM textures in solids.  
        (a,b) Illustrations of SAM (a) and OAM (b) of an electron (yellow spheres). Blue and red arrows represent angular momenta ($\vec{S}$ and $\vec{L}$) and their corresponding magnetic moments ($\vec{m}_{s}$ and $\vec{m}_{L}$), respectively.
        (c-f) Various $\vec{k}$-space OAM textures: helical OAM texture (c); radial OAM texture (d); OAM vortex line (e); and OAM monopole (f). 
        (g-j) $\vec{k}$-dependent SAM textures in magnets with multipolar symmetry form factors: \textit{s}- (g); \textit{p}- (h); \textit{d}- (i); and \textit{g}-wave SAM textures (j). Blue and red colors denote spin-up and spin-down polarizations, respectively.
        }
        \label{fig:1}
\end{figure*}

The spin angular momentum (SAM, $\vec{S}$) and orbital angular momentum (OAM, $\vec{L}$) of the electron share a close conceptual analogy \cite{hirst_microscopic_1997}. The SAM, a fundamental quantum property of the electron, gives rise to a spin magnetic moment described by the relation $\vec{m}_{s}$ = $-eg_{S}\vec{S}/{2m_{e}}$, where $e$ is the elementary charge, $g_{S}$ ($\approx$ 2) is the spin $g$-factor, and $m_{e}$ is the electron rest mass [Fig.~\ref{fig:1}(a)]. Similarly, an electron can exhibit OAM when it physically rotates in real space [Fig.~\ref{fig:1}(b)]. This circular motion of electron generates orbital magnetic moment given by $\vec{m}_{L}$ = $-eg_{L}\vec{L}/{2m_{e}}$, where $g_{L}$ (= 1) is an orbital $g$-factor. It is important to note that both the SAM and OAM contribute to the net magnetization of materials \cite{scott_review_1962,reck_orbital_1969,hirst_microscopic_1997}. Despite these similarities, spin-based functionalities$-$such as information processing using SAM, commonly referred to as spintronics$-$have seen remarkable progress, whereas OAM has only recently begun to attract attention for its significance. This discrepancy largely stems from the long-standing assumption that OAM is quenched by the crystal field in crystalline solids \cite{kittel_introduction_2005}. However, recent theoretical and experimental advances have revealed mechanisms that can generate OAM in electronic states, both dynamically and statically, challenging this long-held view \cite{bernevig_orbitronics_2005,park_orbital-angular-momentum_2011,sunko_maximal_2017,go_intrinsic_2018,choi_observation_2023,ding_observation_2022,el_hamdi_observation_2023,burgos_atencia_orbital_2024}.

Driven by the growing interest in OAM in crystalline systems, various forms of momentum space ($\vec{k}$-space) OAM textures have recently been explored. Representative examples include the Rashba-type surface states in the two-dimensional electron gas systems and the topologically protected surface states of topological insulators, both of which host helical OAM textures and the associated orbital Edelstein effect [Fig.~\ref{fig:1}(c)] \cite{park_orbital-angular-momentum_2011,park_chiral_2012,kim_microscopic_2013,unzelmann_orbital-driven_2020-1,ding_observation_2022}. In addition, theoretical predictions have suggested the emergence of radial OAM textures in elemental selenium crystals [Fig.~\ref{fig:1}(d)] \cite{kim_optoelectronic_2023}. Such OAM textures defined in two-dimensional $\vec{k}$-space have further been extended into three-dimensions, giving rise to OAM vortex lines [Fig.~\ref{fig:1}(e)] \cite{figgemeier_imaging_2025} and OAM monopoles [Fig.~\ref{fig:1}(f)] \cite{yang_monopole-like_2023,yen_controllable_2024}. Thus, the discovery of novel $\vec{k}$-space OAM textures is promoting the advancement of OAM physics and may open the door for device applications that leverage the orbital degree of freedom, a paradigm often referred to as orbitronics.

It is worth noting that these $\vec{k}$-space OAM textures closely mirror their SAM counterparts. For example, the helical OAM texture exhibiting orbital-momentum locking is analogous to the spin-momentum locking with helical SAM texture, while the radial OAM texture finds its analogue in the radial SAM texture \cite{kim_microscopic_2013,sakano_radial_2020,gatti_radial_2020,kim_optoelectronic_2023}. Such a one-to-one correspondence between $\vec{k}$-space OAM and SAM textures naturally raises the outstanding question: can OAM textures with multipolar symmetry form factors$-$analogous to the pairing symmetry form factors in superconductors and to the $\vec{k}$-dependent SAM textures recently realized in unconventional magnets$-$also exist? \cite{smejkal_beyond_2022,smejkal_emerging_2022,krempasky_altermagnetic_2024,lee_broken_2024,zhu_observation_2024,jiang_metallic_2025-1,song_altermagnets_2025,song_electrical_2025,ezawa_third-order_2025}. In this context, exploring OAM textures that mimic the experimentally verified \textit{d}- and \textit{g}-wave SAM textures observed in altermagnets [Fig.~\ref{fig:1}(i) and~\ref{fig:1}(j)] \cite{krempasky_altermagnetic_2024,jiang_metallic_2025-1} as well as the \textit{p}-wave SAM textures of certain noncollinear magnets [Fig. ~\ref{fig:1}(h)] \cite{song_electrical_2025}, is an important frontier and an exciting challenge for advancing OAM physics.

Here, we report on the $\vec{k}$-space imaging of a \textit{p}-wave OAM texture in the 1D chiral crystal (TaSe$_{4}$)$_{2}$I. Using circular dichroism angle-resolved photoemission spectroscopy (CD-ARPES), we directly visualize the dipolar $\vec{k}$-space OAM texture, which displays a characteristic \textit{p}-wave symmetry form factor. We further find that the polarity of the odd-parity OAM pattern reverses between two distinct enantiomers (two mirror counterparts for chiral objects) of (TaSe$_{4}$)$_{2}$I, as expected on symmetry grounds. This finding indicates that the \textit{p}-wave OAM texture is controllable via structural chirality. Complementary spin-resolved ARPES (SARPES) measurements reveal negligible $\vec{k}$-space spin splitting in (TaSe$_{4}$)$_{2}$I, consistent with first-principles calculations. These results strongly suggest that OAM dominates the low-energy electronic properties of (TaSe$_{4}$)$_{2}$I, underscoring its potential as an ideal platform for spinless orbitronics applications. Furthermore, our work provides a pathway toward realizing pure OAM phenomena with minimal SAM contributions, while also laying the fundamental groundwork for realizing multipolar OAM textures, orbital counterparts of spin textures with higher-order form factors in unconventional magnets \cite{smejkal_beyond_2022,smejkal_emerging_2022,lee_broken_2024,krempasky_altermagnetic_2024,jiang_metallic_2025-1,song_altermagnets_2025,ezawa_third-order_2025}. 

\begin{figure*}[htbp!]
	\includegraphics[width=16.4cm]{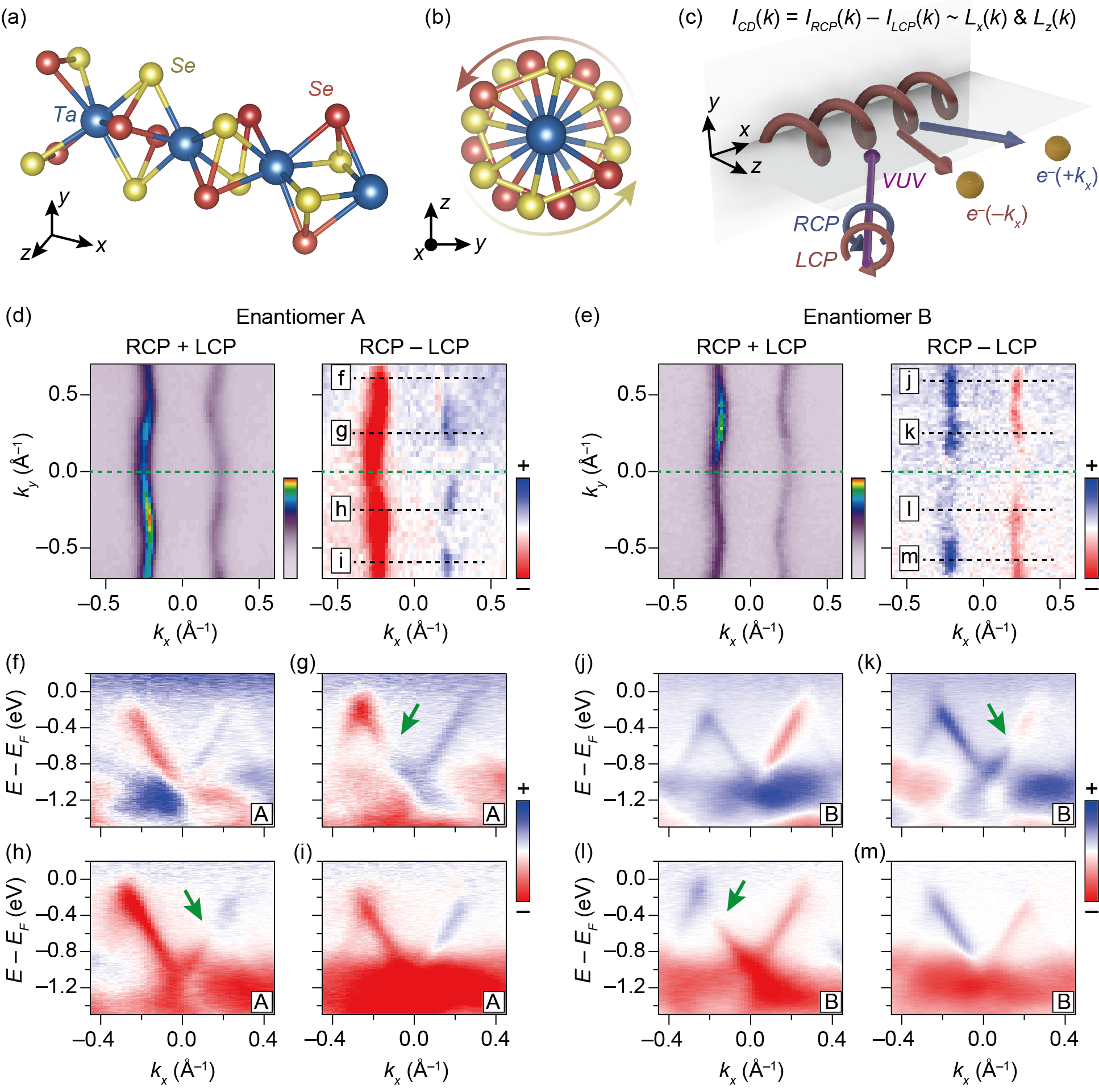}
	\caption{  
        CD-ARPES of (TaSe$_{4}$)$_{2}$I.
		(a) 1D chiral chain structure composed of Se (red and yellow spheres) and Ta (blue sphere) atoms.
        (b) \textit{x}-axis view of the crystal structure of (TaSe$_{4}$)$_{2}$I. Transparency of the sphere reflects the atomic height along \textit{x}-direction.
        (c) Experimental geometry of CD-ARPES.
        (d,e) Constant energy contour maps of enantiomer A (d) and enantiomer B (e) obtained at \textit{E} = $-$0.25 eV with 75 eV photon energy. The left and right panels of (d) and (e) show polarization-integrated and CD photoelectron intensities, respectively. The green dashed lines indicate the plane of incidence of the experimental geometry.
        (f-m) CD-ARPES spectra of enantiomer A (f-i) and enantiomer B (j-m), extracted along the dashed lines marked in (d) and (e). The green arrows denote the node where the CD signal changes sign. 
        }
        \label{fig:2}
\end{figure*}

\section{CD-ARPES on (T\lowercase{a}S\lowercase{e}\textsubscript{4})\textsubscript{2}I}
Figure~\ref{fig:2}(a) illustrates the crystal structure of (TaSe$_{4}$)$_{2}$I, which features a highly anisotropic 1D chain composed of Ta and Se atoms aligned along the \textit{x}-axis. Notably, the Se atoms form a helical structure that wraps around the central 1D Ta chain, as illustrated in Figs.~\ref{fig:2}(a) and~\ref{fig:2}(b). The room temperature metallic electronic structure of (TaSe$_{4}$)$_{2}$I in the low energy range, characterized by linear band crossings, contains contributions from both Se 4\textit{p} and Ta 5\textit{d} orbital characters (see Fig. S1 in Supplemental Material) \cite{Sup_TSI}.

To experimentally probe the OAM texture in this system, we carried out CD-ARPES experiments on (TaSe$_{4}$)$_{2}$I single crystals. Recent experimental and theoretical advancements have established that CD-ARPES$-$defined as the difference in ARPES intensity obtained using right- and left-circularly polarized (RCP and LCP) light, $I_{RCP} (k)-I_{LCP} (k)$$-$can effectively probe momentum-space OAM textures \cite{park_orbital-angular-momentum_2011,park_orbital_2012, park_chiral_2012, unzelmann_orbital-driven_2020-1,unzelmann_momentum-space_2021, brinkman_chirality-driven_2024, figgemeier_imaging_2025,oh_interplay_2025}. This technique is sensitive to OAM components aligned parallel to the direction of light propagation \cite{figgemeier_imaging_2025, moser_toy_2023}. Accordingly, when the incident light impinges obliquely on the sample surface within the plane of incidence [corresponding to the \textit{xz}-plane in Fig.~\ref{fig:2}(c)], it enables detection of both \textit{x}- and \textit{z}-components of the OAM ($L_{x}$ and $L_{z}$). Therefore, proper alignment of the sample's crystallographic axes relative to the experimental geometry is essential to isolate intrinsic OAM contributions in the CD-ARPES signal. To this end, we precisely oriented the chain axis of (TaSe$_{4}$)$_{2}$I along the \textit{x}-axis of the experimental geometry, as depicted in Fig.~\ref{fig:2}(c), to facilitate detection of the $L_{x}$ component which is oriented parallel to the chain direction. This contrasts with previous CD-ARPES studies on (TaSe$_{4}$)$_{2}$I, where the chain axis was aligned perpendicular to the plane of incidence \cite{kim_kramers-weyl_2021, yi_surface_2021}. Additionally, in line with standard CD-ARPES methodology, the experimental geometry was carefully configured such that the surface normal of the cleaved crystal lay within the plane of incidence \cite{wang_observation_2011,cho_experimental_2018}. Note that all ARPES measurements were conducted at room temperature (300 K) to avoid the complex lattice distortions associated with the charge density wave transition of (TaSe$_{4}$)$_{2}$I ($T_{CDW}\sim$ 275 K), allowing us to focus on the intrinsic properties stemming from its pristine chiral structure.

Figure~\ref{fig:2}(d) and~\ref{fig:2}(e) show the measured constant energy contours at $E$ = $-$0.25 eV near the first Brillouin zone for the A and B enantiomers of (TaSe$_{4}$)$_{2}$I, obtained using 75 eV photon energy. As shown in the polarization-integrated ARPES spectra [$I_{RCP}+I_{LCP}$, in the left panels of the Fig.~\ref{fig:2}(d) and~\ref{fig:2}(e)], (TaSe$_{4}$)$_{2}$I displays a strongly anisotropic electronic band structure characterized by open Fermi contours, highlighting the 1D nature of its low-energy electronic states, in agreement with previous ARPES studies \cite{yi_surface_2021,tournier-colletta_electronic_2013-1,lin_unconventional_2024}. It is worth noting that the polarization-integrated maps of enantiomers A and B exhibit distinct spectral intensity distributions. While both enantiomers display stronger intensity in the band at negative $k_{x}$, enantiomer A shows enhanced intensity in the negative $k_{x}$ and $k_{y}$ region [third quadrant in Fig.~\ref{fig:2}(d)], whereas enantiomer B exhibits stronger intensity in the negative $k_{x}$ and positive $k_{y}$ region [second quadrant in Fig.~\ref{fig:2}(e)]. These enantiomer-specific features in the photoelectron maps highlight the chirality-dependent electronic properties of (TaSe$_{4}$)$_{2}$I. We note that two bands forming the linear crossing are not well resolved in the constant energy contour data at $E = -0.25$ eV, but they become clearly distinguishable at higher binding energies [see Fig. S2 in Supplemental Material] \cite{Sup_TSI}. However, the spectral weight of the outer bands is significantly suppressed due to matrix-element effects. Therefore, the spectral weight observed in Figs.~\ref{fig:2}(d) and~\ref{fig:2}(e) is dominated by the inner bands. 

A particularly intriguing feature was observed in the constant energy contours obtained from the CD-ARPES measurements. The CD maps [right panels of the Figs.~\ref{fig:2}(d) and~\ref{fig:2}(e)], which are likewise predominantly contributed by the CD signal of the inner bands, clearly show that enantiomers A and B exhibit opposite odd-parity CD signals, reflecting their distinct structural chirality. As further illustrated in Fig.~\ref{fig:2}(f)-(i), the electronic band of enantiomer A exhibits negative CD (marked with the red color scale) in the negative $k_{x}$ region at low energies ($E-E_{F}\geq -0.8$ eV), largely independent of $k_{y}$. In contrast, the band in the positive $k_{x}$ region within the same energy window displays positive CD [Figs.~\ref{fig:2}(f)-(i)]. On the other hand, odd-parity CD for enantiomer B shows the opposite behavior, with positive CD in the $-k_{x}$ region and negative CD in the $+k_{x}$ region [Figs.~\ref{fig:2}(j)-(m)], confirming the enantiomer-dependent CD response. Based on these CD-ARPES experiemtns, we can experimentally infer not only that the low-energy bands located in the $-k_{x}$ and $+k_{x}$ regions possess OAM components oriented in opposite directions, but also that the polarity of the OAM texture can be controlled through structural chirality.

Not only for the sign reversal of CD between enantiomers A and B, a global inversion of the overall momentum distribution of the CD is also observed. Comparing the CD-ARPES spectra of enantiomers A and B at $k_{y}$ = +0.25 Å$^{-1}$ [Figs.~\ref{fig:2}(g) and~\ref{fig:2}(k)], the CD node indicated by green arrows, where the sign of the CD signal reverses, appears at $k_{x}$ = $-$0.2 Å$^{-1}$, \textit{E} = $-$0.6 eV for enantiomer A, whereas it shifts to $k_{x}$ = +0.2 Å$^{-1}$ for enantiomer B at the same energy. A similar inversion of the CD-ARPES spectra is also observed at $k_{y}$ = $-$0.25 Å$^{-1}$ [Figs.~\ref{fig:2}(h) and~\ref{fig:2}(i)]. This behavior originates from the fact that enantiomers A and B are mirror counterparts of each other [see Fig. S3 in Supplemental Material] \cite{Sup_TSI}. As structural mirror images, their CD-ARPES spectra are correspondingly related through mirror symmetry. Applying a mirror operation along the $x$-direction ($M_{x}$) to the CD-ARPES spectrum in Fig.~\ref{fig:2}(g) (the $+k_{y}$ region) reverses the sign of $k_{x}$ while leaving the sign of $L_{x}$ and $k_{y}$ unchanged, since $\vec{k}$ is a polar vector and $\vec{L}$ is an axial vector \cite{rodriguez-carvajal_symmetry_2012}. This transformation reproduces the CD-ARPES spectrum shown in Fig. 2k [see Fig. S3(b) in Supplemental Material] \cite{Sup_TSI}. The same mirror symmetry relationship holds between Figs.~\ref{fig:2}(h) and~\ref{fig:2}(l). Furthermore, applying a mirror operation along the $y$-direction ($M_{y}$) to the CD-ARPES data in Figs.~\ref{fig:2}(g) and~\ref{fig:2}(h) leaves $k_{x}$ unchanged but reverses the signs of $k_{y}$ and $L_{x}$, thereby reproducing Figs.~\ref{fig:2}(l) and~\ref{fig:2}(k), respectively [see Fig. S3(c) in Supplemental Material] \cite{Sup_TSI}.

We note that, although the $L_{z}$ component can in principle contribute to the CD-ARPES spectra under our experimental geometry, the symmetry relations discussed above indicate that its contribution to the observed CD-ARPES spectra is negligible. When the OAM aligns along the $z$-direction, it undergoes a sign reversal under both $M_{x}$ and $M_{y}$, in contrast to $L_{x}$ [see Fig. S4 in Supplemental Material] \cite{Sup_TSI}. Consequently, the sign of $L_{z}$ would remain unchanged under $M_{x}$, which is incompatible with the observation that Figs.~\ref{fig:2}(g) and~\ref{fig:2}(k) [and likewise Figs.~\ref{fig:2}(h) and~\ref{fig:2}(l)] are related as $M_{x}$ mirror counterparts. Together, these symmetry characteristics demonstrate that the observed CD-ARPES spectra predominantly reflect $L_{x}$, and that these unique CD-ARPES features in (TaSe$4$)$_{2}$I arise directly from its structural chirality.

\begin{figure*}[htbp!]
	\includegraphics[width=16cm]{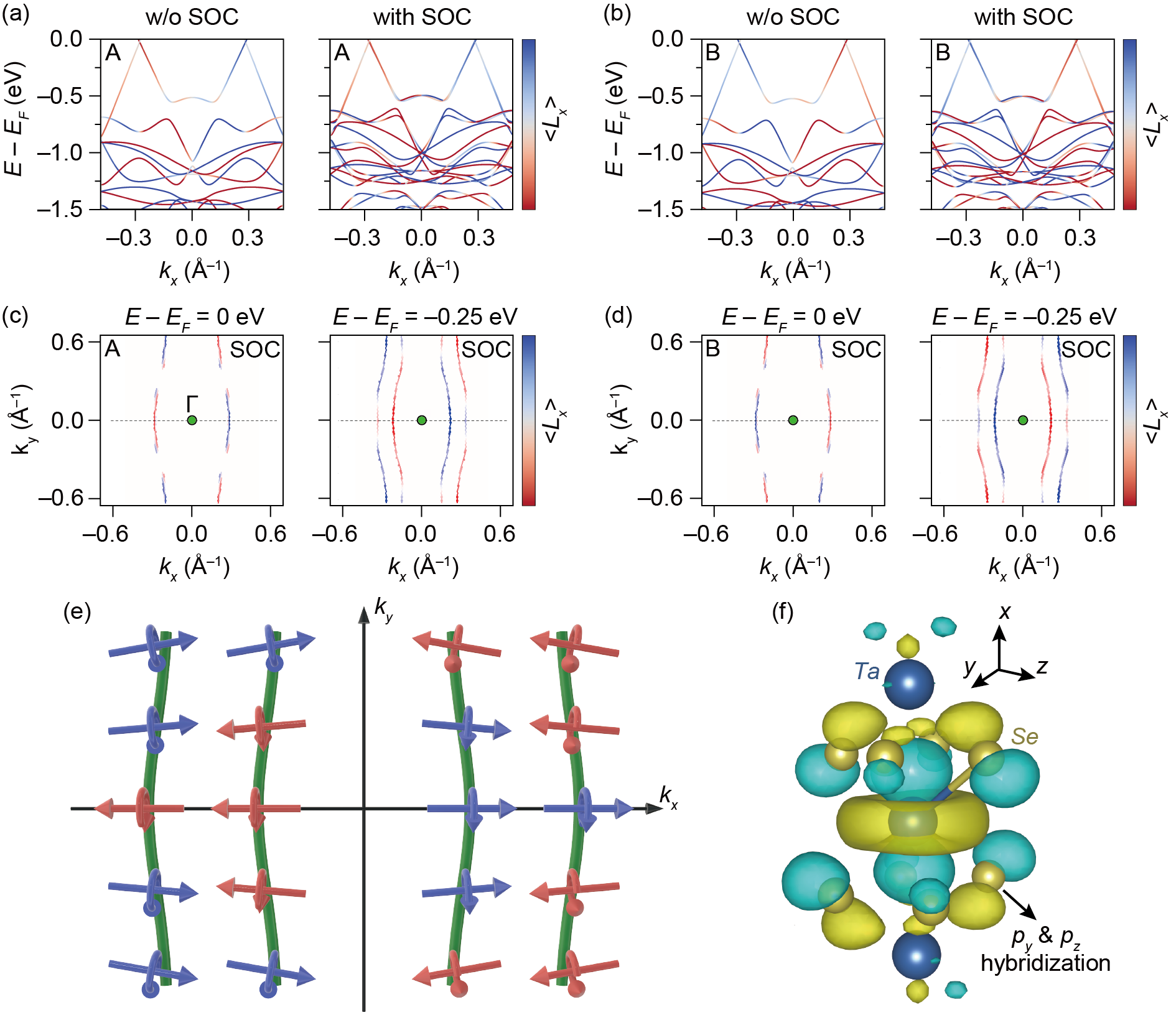}
	\caption{  
		\textit{p}-wave OAM texture in (TaSe$_{4}$)$_{2}$I.
        (a,b) \textit{x}-component of OAM ($L_{x}$) textures in the energy-momentum dispersion of enantiomer A (a), and B (b), extracted at $k_{y}$ = 0 \AA$^{-1}$. Left and right panels in (a) and (b) show results without and with SOC, respectively. 
        (c,d) $L_{x}$ textures of enantiomer A (c) and B (d) in the $k_{x}$-$k_{y}$ plane. Left and right panels illustrate constant energy contours at $E-E_{F}$ = 0 eV and $-$0.25 eV, respectively.
        (e) Schematic illustration of $\vec{k}$-space OAM texture. Green solid curves represent the electronic bands of (TaSe$_{4}$)$_{2}$I in $k_{x}$-$k_{y}$ plane. Red and blue arrows indicate negative and positive $L_{x}$, respectively. The momentum spacing between the inner and outer bands is exaggerated for clarity.
        (f) Real-space Wannier function corresponding to the low-energy electronic states. Blue and yellow spheres correspond to Ta and Se atoms, respectively. Cyan and yellow iso-surfaces represent positive and negative values of the Wannier funtion, respectively. 
        }
        \label{fig:3}
\end{figure*}

It is worth emphasizing that the observed CD-ARPES signals are unlikely to arise from extrinsic geometrical effects \cite{cho_experimental_2018,erhardt_bias-free_2024,brinkman_chirality-driven_2024}, which are commonly known to influence CD-ARPES measurements. This conclusion is supported by several key observations. First, in the low-energy region within the first Brillouin zone ($|k_{x}|$ < 0.5 Å$^{-1}$), the CD signal shows little to no sign reversal across the $k_{y}$ axis (which is perpendicular to the plane of incidence in our experimental geometry), a behavior inconsistent with extrinsic geometrical CD. When the plane of incidence is aligned with the \textit{xz}-plane, obliquely incident circularly polarized light inherently breaks the horizontal mirror symmetry ($M_{y}$) with respect to the \textit{xz}-plane. As a result, if the surface normal of the sample lies within the plane of incidence, the ARPES spectra taken along the $k_{y}$ direction generally become asymmetric with respect to $k_{y}$ = 0 \AA$^{-1}$ for both RCP and LCP light, yielding $I_{RCP}(k_{x}, +k_{y}) \neq I_{RCP}(k_{x}, -k_{y})$ [see Fig. S5 in Supplemental Material] \cite{Sup_TSI}. Moreover, since RCP and LCP light are mirror counterparts under $M_{y}$, the relation $I_{RCP}(k_{x}, \pm k_{y}) = I_{LCP}(k_{x}, \mp k_{y})$ holds. This leads to an extrinsic geometrical CD signal that is, by symmetry, an odd function with respect to $k_{y}$ = 0 plane, typically producing a horizontal node where the CD signal vanishes at $k_{y}$ = 0 \AA$^{-1}$, and is independent of the material's intrinsic electronic properties. However, contrary to this notion, enantiomers A and B of (TaSe$_{4}$)$_{2}$I show no sign reversal of the CD signal across $k_{y}$ = 0 \AA$^{-1}$ in either the positive or negative $k_{x}$ regions, as evident in Figs.~\ref{fig:2}(d)-(m). This strongly suggests that the CD signals in these momentum and energy ranges are governed primarily by intrinsic CD contributions, rather than extrinsic effects. Secondly, the observed odd-parity CD-ARPES spectra is reversible between the two enantiomers, whereas the extrinsic geometrical CD is enantiomer-independent. The CD signals exhibiting sign reversal across $k_{y}=0$ \AA$^{-1}$, a hallmark of extrinsic geometrical effects, appear only at higher binding energies and far from the Brillouin zone center [see Fig. S6 in Supplemental Material] \cite{Sup_TSI}. Crucially, these features are present in both enantiomers and are independent of their structural chirality, further confirming their extrinsic origin. Taken together, these observations collectively support the conclusion that the enantiomer-dependent odd-parity CD-ARPES data present in Fig.~\ref{fig:2} faithfully reflects intrinsic CD contributions arising from inherent electronic states intertwined with the chiral lattice structure of (TaSe$_{4}$)$_{2}$I. We also note that while extrinsic geometrical CD contributions are negligible in the low-energy spectra measured using 75 eV photon energy, they become more pronounced at higher photon energies [see Fig. S7 in Supplemental Material] \cite{Sup_TSI}. In addition, photon energy-dependent CD sign reversals, commonly observed in CD-ARPES, are also detected. The photon energy dependence in CD-ARPES spectra may originate from final-state effect and inter-atomic interference \cite{erhardt_bias-free_2024,yen_controllable_2024,sidilkover_reexamining_2025,brinkman_chirality-driven_2024}. We further note that the observed odd-parity momentum distribution of the CD can be attributed to the OAM character of the initial electronic states, rather than to optical activity effects [see Supplementary Section SVII] \cite{Sup_TSI}. Therefore, we conclude that the measured CD-ARPES spectra are a direct manifestation of the OAM texture of the initial electronic states of (TaSe$_{4}$)$_{2}$I.

\section{\textit{\lowercase{p}}-wave OAM texture in (T\lowercase{a}S\lowercase{e}\textsubscript{4})\textsubscript{2}I}

To obtain deeper insight into the observed angular distribution of the CD signal in the low-energy regime, we performed first-principles calculations to evaluate the orbital degree of freedom in (TaSe$_{4}$)$_{2}$I. Fig.~\ref{fig:3}(a) presents the calculated energy-momentum dispersion of enantiomer A projected onto $L_{x}$. In the non-relativistic limit (without spin-orbit coupling, SOC), shown in the left panel of Fig.~\ref{fig:3}(a), enantiomer A exhibits linearly dispersing low-energy bands with $+L_{x}$ in the $+k_{x}$ region, while the sign of $L_{x}$ reverses in the $-k_{x}$ region, demonstrating an odd-parity $L_{x}$ texture. Interestingly, even in the relativistic limit with SOC, the low-energy bands exhibit a small energy splitting (< 5 meV), and both split branches retain the same sign of $L_{x}$ [right panel of Fig.~\ref{fig:3}(a)]. This result indicates that the low-energy OAM textures are almost identical in the non-relativistic and relativistic limits of (TaSe$_{4}$)$_{2}$I. Moreover, enantiomer B hosts an OAM texture with the opposite sign of $L_{x}$ compared to that of enantiomer A, as shown in Fig.~\ref{fig:3}(b). Notably, the $\vec{k}$-dependent OAM texture at $k_{y}$ = 0 \AA$^{-1}$ consists solely of the $L_{x}$ component, while the perpendicular components $L_{y}$ and $L_{z}$ are absent [see Fig. S9 in Supplemental Material] \cite{Sup_TSI}.

We further investigate the $\vec{k}$-space OAM texture away from $k_{y}$ = 0 \AA$^{-1}$. Fig.~\ref{fig:3}(c) and~\ref{fig:3}(d) display constant energy contours in the $k_{x}$-$k_{y}$ plane projected onto $L_{x}$. At the Fermi level of enantiomer A [left panel of Fig.~\ref{fig:3}(c)], the OAM texture for positive $k_{x}$ clearly shows positive $L_{x}$ near $k_{y}$ = 0 \AA$^{-1}$, as both the inner (smaller $k_{x}$) and outer (larger $k_{x}$) bands share the same sign of $L_{x}$. In contrast, the bands at negative $k_{x}$ exhibit the opposite sign of $L_{x}$, demonstrating a dipolar OAM texture with \textit{p}-wave symmetry form factor. Near the Brillouin zone boundary, the OAM undergoes a sign reversal, in reasonable agreement with the CD-ARPES measurements [see Fig. S5 in Supplemental Material] \cite{Sup_TSI}. This \textit{p}-wave OAM texture becomes even more pronounced at higher binding energy ($E-E_{F}$ = $-$0.25 eV), as shown in the right panel of Fig.~\ref{fig:3}(c). The polarity inversion of the \textit{p}-wave OAM texture in enantiomer B is also more distinctly captured at this binding energy [Fig. 3(d)]. To obtain a complete picture, we thoroughly examine the full OAM texture of (TaSe$_{4}$)$_{2}$I by further evaluating the $L_{y}$ and $L_{z}$ components, as illustrated in Fig.~\ref{fig:3}(e) [see Fig. S10 Supplementary] \cite{Sup_TSI}. Although the low-energy band structure of (TaSe$_{4}$)$_{2}$I contains a weak $L_{y}$ component (while $L_{z}$ remains absent), the overall OAM structure still predominantly exhibits a \textit{p}-wave symmetry form factor [Fig.~\ref{fig:1}(h)]. These first-principles calculations are in good agreement with the CD patterns observed in the low-energy ARPES spectra (Fig.~\ref{fig:2}), reinforcing the identification of a \textit{p}-wave OAM texture in (TaSe$_{4}$)$_{2}$I.

It is important to emphasize that the characteristic $p$-wave OAM texture observed in (TaSe$_{4}$)$_{2}$I is distinct from the texture obtained by transforming a radial OAM texture$-$defined on a closed Fermi contour$-$to the 1D limit where the contour becomes open [see Fig. S11 in Supplemental Material] \cite{Sup_TSI}. For example, in the simple radial OAM case, the OAM vectors in the $+k_{x}$ and $+k_{y}$ regions (the first quadrant in $k_{x}$-$k_{y}$ plane) have the same signs for $L_{x}$ and $L_{y}$. In contrast, in (TaSe$_{4}$)$_{2}$I, the OAM vectors in the same $\vec{k}$ region exhibit opposite signs for $L_{x}$ and $L_{y}$ [see Fig. S10 in Supplemental Material] \cite{Sup_TSI}. Moreover, approximating the OAM texture of (TaSe$_{4}$)$_{2}$I as a 1D limit of a simple radial configuration fails to account for the sign reversal observed along its constant energy contours [Fig. 3(c) and 3(d)]. Therefore, these observations indicate that the $p$-wave OAM texture can be regarded as an inherent multipolar OAM characteristic of (TaSe$_{4}$)$_{2}$I.

\begin{figure*}[t!]
	\includegraphics[width=15cm]{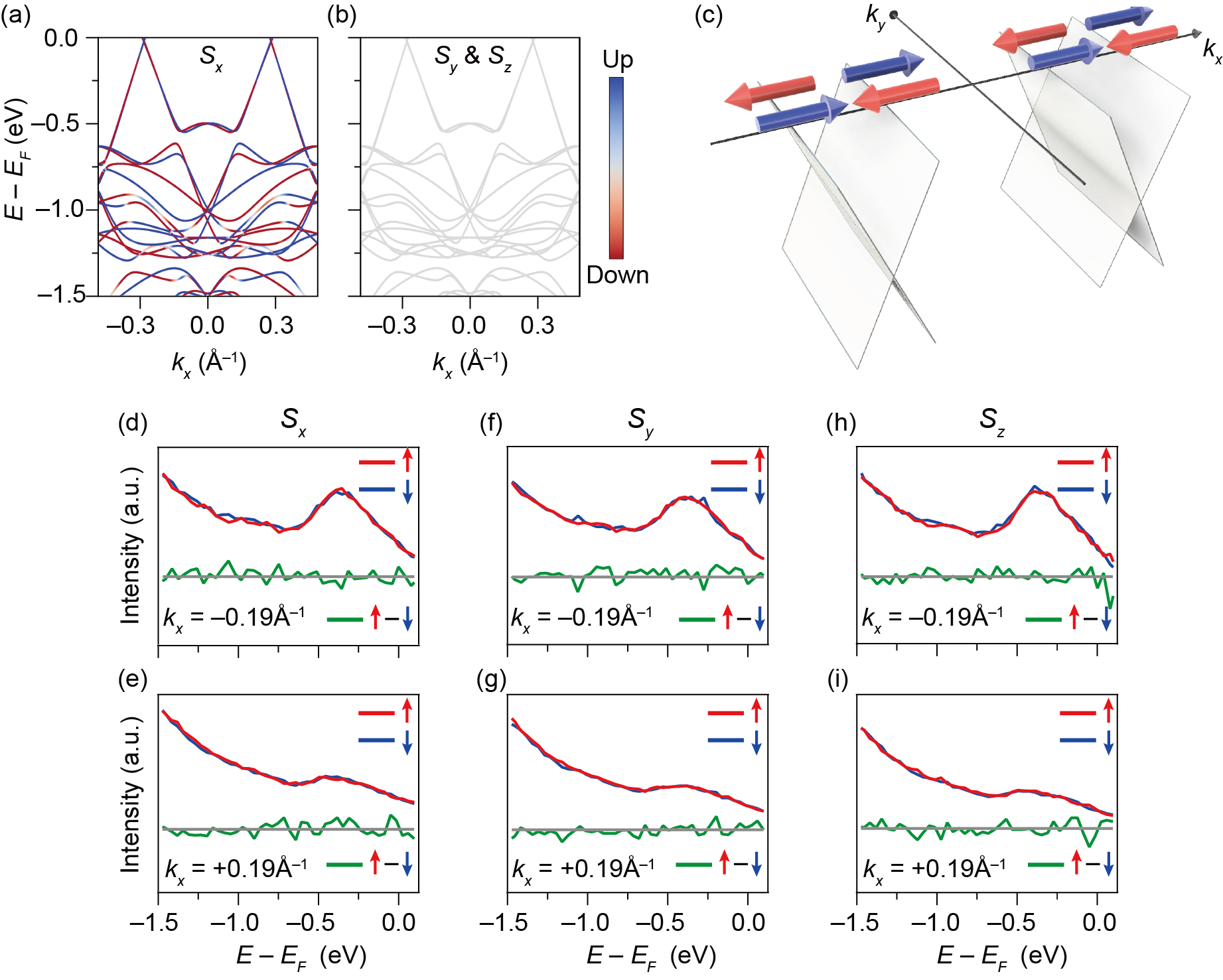}
	\caption{  
		SAM texture in (TaSe$_{4}$)$_{2}$I.
        (a,b) $\vec{k}$-space SAM texture of (TaSe$_{4}$)$_{2}$I at $k_{y}$ = 0 \AA$^{-1}$, projected onto the \textit{x}-component of SAM, $S_{x}$, (a) and \textit{y}- and \textit{z}-components, $S_{y}$ and $S_{z}$, (b). 
        (c) Schematic illustration of the $\vec{k}$-space SAM texture in the low-energy electronic states of TaSe$_{4}$)$_{2}$I. Grey planes indicate the linearly dispersing bands. Red and blue arrows correspond to positive and negative $S_{x}$, respectively.
        (d-i) Spin-resolved ARPES (SARPES) spectra of (TaSe$_{4}$)$_{2}$I. (d) and (e) show spin-resolved energy distribution curves (SEDCs) projected onto $S_{x}$, collected at $\mp$0.19 \AA$^{-1}$, respectively. (f) and (g), as well as (h) and (i), display the same SEDCs, but projected onto the $S_{y}$ and $S_{z}$ spin components, respectively. Red and blue indicate spin-up and spin-down components, respectively, while green curves show the differences between the spin-up and spin-down channels.
        }
        \label{fig:4}
\end{figure*}

To shed light onto the microscopic origin of the finite OAM, we analyze the linearly dispersing low-energy bands within the framework of a minimal tight-binding model [see Fig. S1 in Supplemental Material] \cite{Sup_TSI}. Specifically, we project the first-principles electronic band structure onto a set of maximally localized Wannier functions derived from Ta-$d_{x^2}$ orbitals, i.e., the conventional $d_{z^2}$ orbitals aligned along the $x$-direction of the chiral chains. As shown in Fig.~\ref{fig:3}(f), the resulting Wannier functions, while dominated by $d_{x^2}$ character, also exhibit substantial $p$-like contributions from the Se atoms. Importantly, this $p$ orbital character is inherited from the Wannier functions in order to faithfully reproduce and interpolate the original DFT bandstructure. This is consistent with the orbitally resolved DFT band structure in Fig. S1 of Supplemental Material, which highlights a sizable Se contribution to the relevant bands. Consequently, because the $d_{x^2}$ component of the Wannier function does not contribute to the local OAM, the finite $L_{x}$ polarization of the bands originates from the $p_y$ and $p_z$ orbitals within the plane perpendicular to the chiral chains. They indeed act as orbital polarizers of the Wannier functions.

\section{{$\vec{k}$}-space SAM texture in (T\lowercase{a}S\lowercase{e}\textsubscript{4})\textsubscript{2}I}

We further examine the $\vec{k}$-space SAM texture in the electronic structure of (TaSe$_{4}$)$_{2}$I, given that electronic bands hosting OAM textures typically acquire concomitant SAM through SOC. As shown in Figs.~\ref{fig:4}(a) and~\ref{fig:4}(b), the calculated spin textures exhibit a finite $S_{x}$ component, mirroring the behavior observed in the OAM sector, while the $S_{y}$ and $S_{z}$ components are absent. Moreover, the calculated spin polarizations of enantiomer A and B display opposite signs of $S_{x}$, consistent with the corresponding OAM textures [see Fig. S12 in Supplemental Material] \cite{Sup_TSI}. Interestingly, as discussed above, the SOC-induced band splitting near the Fermi level is negligibly small, and the two split branches carry opposite SAM, resulting in an almost vanishing net spin polarization at the Fermi surface. Since this spin-splitting energy scale (< 5 meV) is far below the thermal energy at room temperature ($\sim 25$ meV), the spin split bands are thermally activated, implying that the SAM contribution to the low-energy electronic properties of (TaSe$_{4}$)$_{2}$I is negligible at room temperature. Accordingly, the effective $\vec{k}$-space SAM texture can be represented as shown in Fig.~\ref{fig:4}(c). Notably, while the spin-split bands possess opposite SAM signs, they retain the same sign of OAM (Fig.~\ref{fig:3}), indicating that only OAM polarization persists at the Fermi surface. These unique OAM and SAM textures suggest that $\vec{L}$ serves as the good quantum number governing the low-energy electronic states, with SAM being secondarily locked to the OAM through weak SOC \cite{kim_microscopic_2013}. In addition, the dominant energy scale of inversion-symmetry breaking in (TaSe$_{4}$)$_{2}$I, arising from its chiral structure and exceeding that of SOC, is likely a key factor in these SAM and OAM structures \cite{sunko_maximal_2017}. 

The SAM textures in (TaSe$_{4}$)$_{2}$I obtained from first-principles calculations show reasonably good agreement with our SARPES measurements conducted at room temperature. As shown in Fig.~\ref{fig:4}(d) and~\ref{fig:4}(e), spin-resolved energy distribution curves collected at $k_{x}=\pm 0.19$ Å$^{-1}$, where the ARPES spectral weight is most pronounced, exhibit no detectable $S_{x}$ polarization, as the tiny spin splitting is intrisically obscured by the limited energy resolution of the SARPES technique and by the thermal broadening at room temperature. These results are nevertheless consistent with the presence of sub-resolution spin splitting projected onto $S_{x}$. Moreover, no discernible spin polarization is observed in $S_{y}$ and $S_{z}$ [Figs.~\ref{fig:4}(f)-~\ref{fig:4}(i)], in line with the first-principles results [Fig.~\ref{fig:4}(b)]. The absence of measurable spin polarization near the Fermi level provides direct experimental evidence that OAM, rather than SAM, has a leading role in the low-energy electronic properties of (TaSe$_{4}$)$_{2}$I. Therefore, (TaSe$_{4}$)$_{2}$I offers a significant advantage for realizing spinless orbitronics applications.

\section{Discussion}
Through a systematic spectroscopic study on (TaSe$_{4}$)$_{2}$I, we find a dipolar OAM texture characterized by a \textit{p}-wave symmetry form factor, with a negligible SAM contribution to the low-energy physics. This makes (TaSe$_{4}$)$_{2}$I a promising material platform for future spinless orbitronics \cite{jo_gigantic_2018,choi_observation_2023}. Furthermore, we demonstrate that the polarity of \textit{p}-wave OAM texture is controllable through structural chirality, reminiscent of the tunable OAM monopoles observed in 3D chiral topological semimetal \cite{yen_controllable_2024}. This observation can be regarded as the orbital counterpart of \textit{p}-wave SAM texture, recently realized in helimagnetic state of NiI$_{2}$ \cite{song_electrical_2025}. From a symmetry perspective, (TaSe$_{4}$)$_{2}$I and NiI$_{2}$, which exhibit \textit{p}-wave OAM and SAM, respectively, share a profound similarity. In the former, inversion symmetry is broken by the chiral crystal structure, whereas in the latter it is broken by the chiral magnetic order. In addition, the polarity of \textit{p}-wave OAM in (TaSe$_{4}$)$_{2}$I and \textit{p}-wave SAM in NiI$_{2}$ can be controlled via chirality in lattice and spin degrees of freedom, respectively. 

By noting that the \textit{p}-wave OAM texture emerging from the helical chain structure can be directly connected to a \textit{p}-wave SAM texture in helimagnet, we further establish the analogy between chiral crystal structures and chiral magnetic orders [see Fig. S13 in Supplemental Material] \cite{Sup_TSI}. In the non-relativistic limit, the helical chain structure hosts a \textit{p}-wave OAM texture. Then, when a sufficiently strong SOC is taken into account, it can give rise to pronounced \textit{p}-wave SAM texture. By pinpointing this microscopic mechanism of coupled OAM and SAM in 1D chiral lattice structures, our study highlights a novel pathway to realize \textit{p}-wave SAM texture driven by structural chirality intertwined with SOC. This approach offers a distinct advantage over unconventional magnets, such as altermagnets, which rely on antiferromagnetic ordering, as the characteristic size of structurally chiral domains is generally much larger than that of antiferromagnetic domains. Harnessing structural chirality to engineer \textit{p}-wave SAM textures thus provides a promising route toward realizing coherent \textit{p}-wave SAM textures over long length scales. Such a strategy could open the door to novel emergent phenomena, including unconventional chiral superconductivity \cite{nakajima_giant_2023}.

Studies of OAM and SAM in (TaSe$_{4}$)$_{2}$I further provide fundamental insight into the innate electronic properties in helical chain structures. Recent studies of quantum materials have proposed that OAM may serve as the microscopic origin of spin transport phenomena such as the spin Hall and Rashba effects \cite{park_orbital_2012,park_orbital-angular-momentum_2011,go_intrinsic_2018,bhowal_intrinsic_2020}. Within this broader context, our findings highlight that OAM could also underlie chirality-induced spin selectivity (CISS), a distinctive spin transport effect in helical chain systems which shows spin current parallel to the chain direction \cite{gohler_spin_2011,naaman_chiral_2019,naaman_chiral_2022}. Specifically, in the absence of SOC, low-energy electronic states carry only OAM, while SOC couples SAM to OAM, generating spin splitting and enabling spin transport phenomena [see Fig. S13 in Supplemental Material] \cite{Sup_TSI}. This framework allows CISS to be reinterpreted as a consequence of more fundamental effect$-$chirality-induced orbital selectivity (CIOS). This connection may also shed light on the long-standing puzzle that the magnitude of CISS depends sensitively on the SOC strength of the electrodes used in measurments \cite{adhikari_interplay_2023,yan_structural_2024}. Taken together, our results provide an important clue toward establishing a correct fundamental understanding of both CISS and CIOS.

\subsection*{Outlook}
In this work, we experimentally and theoretically realize and verify an OAM texture with a \textit{p}-wave symmetry form factor in a crystalline solid. Realizing such an OAM structure establishes a foundation for exploring the orbital analogs of SAM texture with multipolar symmetry form factors in unconventional magnets \cite{smejkal_beyond_2022,smejkal_emerging_2022,lee_broken_2024,krempasky_altermagnetic_2024,jiang_metallic_2025-1,song_altermagnets_2025}. Looking forward, continued experimental and theoretical advances may enable the realization of higher-order multipolar OAM textures, such as \textit{d}-, \textit{f}-, \textit{g}-, and even \textit{i}-wave symmetries, extending beyond the \textit{p}-wave regime \cite{ezawa_third-order_2025}. Moreover, the direct detection of OAM-related phenomena will present the next important challenge. Such effects should be experimentally accessible using state-of-the-art techniques, such as ultra-sensitive magneto-optical Kerr effect measurements capable of resolving Kerr rotation on the order of a few nanoradians \cite{choi_observation_2023}. In this respect, our work provides a fundamental step toward a more complete understanding of OAM-rooted effects in quantum materials.

%%%%%%%%%%%%%

\section*{Acknowledgments}
This work was supported by the Air Force Office of Scientific Research (AFOSR) under grant FA9550-22-1-0432. This research used resources of the Advanced Light Source, which is a DOE Office of Science User Facility under contract no. DE-AC02-05CH11231. D.O. and C.P. contributed equally.

\section*{Appendix: Methods}
\subsection*{1. Single crystal growth}
Single crystals of (TaSe$_{4}$)$_{2}$I were grown using the chemical vapor transport (CVT) method. A mixture of Ta, Se, and I precursors was placed in a quartz tube, which was then sealed under vacuum. The quartz ampule was heated in a two-zone furnace, with the cold and hot zones maintained 400 $^\circ$C and 520 $^\circ$C, respectively. After one week, needle-like (TaSe$_{4}$)$_{2}$I single crystals, a few milimeters in size, were obtained. The stoichiometry of the synthesized crystals was confirmed using Energy-dispersive X-ray spectroscopy (EDX).

\subsection*{2. ARPES measurements}
The CD-ARPES measurements were carried out at Beamline 7.0.2 (MAESTRO) of the Advanced Light Source (ALS). Single crystals of (TaSe$_{4}$)$_{2}$I were mechanically cleaved at room temperature in an ultra-high-vacuum (UHV) ARPES chamber ($\sim$4 $\times$ 10$^{-11}$ torr). The CD-ARPES data were acquired at room temperature using a horizontal analyzer slit. Constant energy contour maps were obtained with a photoelectron deflector, with precise alignment of the experimental geometry to ensure that the plane of incidence include the normal vector of the cleaved (TaSe$_{4}$)$_{2}$I surface. The angle of incidence was set to 54.75$^\circ$.

Spin-resolved ARPES experiments were performed at the Bloch beamline of MAX IV, utilizing very low energy electron diffraction (VLEED) spin detectors. All spin-resolved ARPES data were also collected at room temperature.

\subsection*{3. DFT calculations}
All theoretical calculations were carried out within the framework of ab initio Density functional theory (DFT), implemented in the FPLO (Full-Potential Local-Orbital) package \cite{FPLO1,FPLO2}. The exchange–correlation energy was treated within the generalized gradient approximation (GGA), in the parametrization of Perdew–Burke–Ernzerhof 96 (PBE). Total energies were converged to better than $10^{-8}$ eV, while the accuracy for the self-consistent charge densities was set to $10^{-6}$ eV. For the cases where spin–orbit coupling (SOC) plays a role, fully relativistic calculations were performed using the four-component Dirac formalism implemented in FPLO. The Brillouin zone was sampled using a Monkhorst–Pack k-point mesh of $12\times12\times12$. The relaxed structural parameters were taken from the Materials Project \cite{MP1,MP2,MP3}. 

In order to extract tight-binding (TB) models of the two enantiomers, their Bloch states were projected onto maximally localized Wannier functions using the FPLO Wannier function module. The atomic orbitals chosen as trial functions were Ta-5d, Ta-6s, Se-4p and I-5p. The minimal toy-model based solely on Ta-$d_{z^2}$ orbitals as trial functions was extracted through the Wannier90 package \cite{W90}. The post-processing of the TB models relied on a in-house python code \cite{post_wan}. The Wannier functions were visualized thanks to the Vesta software.

%\bibliography{references}

%

\end{document}